\documentclass{article}

\usepackage[backend=biber,style=numeric]{biblatex}
\usepackage{mathptmx}
\usepackage{titlesec}
\usepackage{titling}
\usepackage{setspace}
\usepackage{indentfirst}
\usepackage{abstract}
\usepackage[margin=1.3in]{geometry}
\usepackage{graphicx}
\usepackage{float}
\usepackage{placeins}

\titleformat{\section}
{\normalfont\large\bfseries}
{\thesection.}{0.5em}{}
\titlespacing{\section}{0pt}{18pt}{6pt}
\makeatletter
\renewcommand\section{\@startsection
  {section}      
  {1}            
  {0pt}          
  {25pt}         
  {5pt}          
  {\normalfont\large\bfseries}
}
\makeatother

\title{AI Advocate: Educational Path to Transform Squads to the Future}

\titlespacing*{\section}{0pt}{0pt}{4pt}

\begin{document}

% ===== TÍTULO =====
\begin{center}
    \rule{\textwidth}{1.5pt} \\[10pt]

    {\LARGE \bfseries 
    AI Advocate: Educational Path to Transform Squads to the Future
    \par}

    \vspace{10pt}
    \rule{\textwidth}{0.5pt} \\[20pt]
\end{center}

% ===== AUTORES =====
\vspace{5pt}

\begin{center}
\begin{minipage}{0.35\textwidth}
    \centering
    \textbf{Carla Soares} \\
    Zup IT Innovation, Brazil \\
    \texttt{carla.soares@zup.com.br}
\end{minipage}
\hspace{2cm}
\begin{minipage}{0.35\textwidth}
    \centering
    \textbf{Gabriel Moreira} \\
    Zup IT Innovation, Brazil \\
    \texttt{gabriel.moreira@zup.com.br}
\end{minipage}
\end{center}

\vspace{12pt}

\begin{center}
\begin{minipage}{0.35\textwidth}
    \centering
    \textbf{Ana Paula Camargo} \\
    Zup IT Innovation, Brazil \\
    \texttt{ana.camargo@zup.com.br}
\end{minipage}
\hspace{2cm}
\begin{minipage}{0.35\textwidth}
    \centering
    \textbf{Fabio Henrique Scacabarozi} \\
    Zup IT Innovation, Brazil \\
    \texttt{fabio.scacabarozi@zup.com.br}
\end{minipage}
\end{center}

\vspace{12pt}

\begin{center}
\begin{minipage}{0.35\textwidth}
    \centering
    \textbf{Nicole Davila} \\
    Zup IT Innovation, Brazil \\
    \texttt{nicole.davila@zup.com.br}
\end{minipage}
\hspace{2cm}
\begin{minipage}{0.35\textwidth}
    \centering
    \textbf{Marselle Silva} \\
    Zup IT Innovation, Brazil \\
    \texttt{marselle.silva@zup.com.br}
\end{minipage}
\end{center}

% \maketitle

\vspace{25pt}

\begin{center}
\begin{minipage}{0.7\textwidth}

\begin{abstract}
\vspace{-4pt} 
This paper analyzes the strategic education process aimed at transitioning traditional software
development squads into hybrid structures, where the focus is defined by collaborative work
between humans and Artificial Intelligence (AI). In a context where human-AI collaboration can
increase productivity by up to 40\% \cite{dellacqua2026}, this study explores how the upskilling
of XPTO professionals—referred to as “AI Advocates”—acts as a catalyst for this cultural and
technical transformation. The objective is to present an experience report on the education and
enablement process of AI Advocates within a private Brazilian technology company, highlighting
key lessons learned and identified challenges.

\textbf{Keywords: }Generative AI; AI-Driven SDLC; Team Topologies; Hybrid Squads; Engineering
Efficiency; Future of Work.
\end{abstract}

\end{minipage}
\end{center}
\vspace{20pt}

\section{Introduction}
The continuous evolution of Generative AI is driving changes across all sectors, from how companies are structured and deliver services to how teams operate. At the core of this transformation is not merely the introduction of new tools, but a broader redefinition of the Future of Work—particularly in the context of software development. According to Harvard Business Review (2023), collaborative integration between humans and Artificial Intelligence (AI) can increase organizational productivity by up to 40\% \cite{dellacqua2025}. However, these gains are not automatic; they require structured training and continuous upskilling to ensure individuals are prepared for these changes.

In modern software development, professionals equipped with AI-based tools are no longer limited to acting as code implementers; instead, they take on the role of orchestrators of complex systems and generative models. For this evolution to scale effectively within organizations, traditional squads must transition into hybrid squads, where humans and AI collaborate side by side. In this emerging paradigm, AI agents take on roles as team members with operational responsibilities, requiring practitioners to develop new technical and non-technical skills.

This paper presents the education and enablement process of AI Advocates, focusing on the transition to hybrid squads: how to train individuals with the potential to act strategically in driving cultural and technical transformation within teams adopting this model. Conducted at Zup Innovation, a Brazilian company in the technology and innovation sector, this initiative resulted in a 6-hour training program delivered to 238 AI Advocates. 

This experience report outlines the main pillars, challenges, and lessons learned throughout the design and execution of the training. The focus was to explore how an AI-based journey enables the practical contextualization of these tools in day-to-day development activities, ensuring that the pursuit of efficiency is consistently aligned with Responsible AI principles—so that innovation remains as ethical as it is productive.

The training program was structured around three core pillars, complemented by two cross-cutting elements. The first pillar, Artificial Intelligence Fundamentals, equipped participants with an understanding of language models, selection criteria, and Responsible AI concepts.

The second pillar, Domain Contextualization, applied Domain-Driven Design principles \cite{evans2003} to the context of AI agents, exploring how to structure domain knowledge for agent consumption. It also covered architecture and context engineering techniques, including Spec-Driven Development \cite{fowler2025}.

The third pillar, Tools and Practice, provided hands-on experience with both market-available and proprietary AI-assisted coding tools.

Across all pillars, the program emphasized preparing participants for the Future of Work, offering a deeper perspective on the evolving role of the engineer, addressing organizational resistance to AI adoption, and supporting the development of impact metrics beyond traditional development time measures.

\section{Project Overview}
The AI Advocates program was conceived at Zup Innovation based on reflections on the essential skills required to operate hybrid squads—that is, teams composed of software engineers and AI agents acting semi-autonomously within the development lifecycle. The primary objective was to develop professionals capable of amplifying the use of AI alongside the company’s clients, disseminating technical knowledge, analyzing existing architectures, and co-creating solutions with development teams.

The term “AI Advocate” was proposed by one of the instructors during the program’s conception phase, replacing initial alternatives such as “AI Software Architect” and “AI Software Developer,” as it more accurately reflects the expected role of knowledge dissemination and influence among participants.

The program is part of a broader organizational strategy that structures AI adoption maturity into progressive levels, ranging from the use of AI as an assistive tool, to hybrid squads with semi-autonomous agents, and ultimately to more advanced agentic orchestration models. The company’s strategic objective is to achieve hybrid squads across the entire organization by 2026, and the AI Advocates program was designed as a key accelerator toward this goal.

\subsection{Organization}
The organization of the program involved assembling a cohort of approximately nine instructors and mentors, designated as trusted advisors by the company’s senior leadership, with contributions from the technical authority function. Prior to delivery, the training plan was reviewed by senior leadership to ensure strategic alignment.

The company’s research team provided foundational insights on the Future of Work, while professionals actively engaged in the external technical community contributed market perspectives. The primary capability identified as critical to enabling the initiative was leadership courage and support, given the experimental nature of the program and the speed required for both training delivery and the scaling of hybrid squads.

During the conception phase, the possibility of engaging external educational institutions to deliver the training was considered. However, time constraints (the program needed to be designed and delivered within a few weeks) made this option unfeasible. This decision ultimately reinforced the focus on internal talent, leveraging the existing knowledge of the organization’s instructors and mentors.

\subsection{Selection Process and Outreach}
Initially, the cohort was planned to include approximately 40 participants, selected by their respective Heads. However, as the program syllabus gained visibility across different leadership levels, interest increased and the audience was expanded. The table below presents the final composition by seniority.

\begin{table}[htbp]
\centering
\caption{Attendance by Professional Level}
\label{tab:attendance_levels}
\renewcommand{\arraystretch}{1.4}
\begin{tabular}{|p{4.5cm}|c|c|c|}
\hline
\textbf{Level} & \textbf{Attended} & \textbf{Did Not Attend} & \textbf{\% of Total} \\
\hline
Junior & 1 & 0 & 0.42\% \\
\hline
Mid-level & 4 & 2 & 1.68\% \\
\hline
Senior & 50 & 7 & 21.01\% \\
\hline
Specialist 1 / Tech Lead & 132 & 18 & 55.46\% \\
\hline
Specialist 2 / Engineering Manager & 51 & 0 & 21.43\% \\
\hline
Total & 238 & 27 & 265 \\
\hline
Attendance Rate & 89.81\% & 10.19\% & -- \\
\hline
\end{tabular}
\end{table}

The predominance of professionals at the Senior and Specialist/Tech Lead levels—who together represented 76.47\% of participants—reflects the intended multiplier profile: individuals with sufficient technical maturity to absorb the content and disseminate it within their teams. The attendance rate of 89.81\% indicates strong engagement with the initiative.

Following the cohort delivery, the content was re-recorded in a closed format for asynchronous distribution across the organization. The program also served as a prelude to subsequent large-scale upskilling initiatives.

\subsection{Structure}
The content was developed over approximately three weeks, totaling six hours delivered across two days and organized into four thematic sessions:

\textbf{Session 1 – Mindset and Strategy.} The first session focused on mindset building and strategic alignment. It covered the company’s purpose, the Future of Work, and ongoing changes in the software engineering landscape, including the distinction between Vibe Coding \cite{karpathy2025} and more structured AI-assisted engineering practices \cite{willison2025}. The session also introduced the company’s strategic vision for AI maturity.

\textbf{Session 2 – AI Fundamentals.} The second session provided a deeper dive into the technical foundations of language models, context and prompt engineering, AI agents and their communication protocols, Responsible AI, and the application of Domain-Driven Design principles \cite{evans2003} in the context of artificial intelligence.

\textbf{Session 3 – Spec-Driven Development and Hands-on.} The third session focused on Spec-Driven Development (SDD)—an approach in which structured specifications, rather than code, serve as the primary development artifact \cite{fowler2025}. It also covered context architecture principles and context window management for agents. The session included practical demonstrations.

\textbf{Session 4 – Tools and Metrics.} The fourth session emphasized hands-on exploration of AI-assisted coding tools, both market-available and internal, as well as discussions on value metrics beyond development time, including code quality and developer experience. The session concluded with the presentation of a roadmap for the program’s continuous evolution.

In addition to the four sessions, follow-up sessions were conducted to address questions and provide deeper practical exploration.

\subsection{Program}
The reflection on the essential skills required for an AI Advocate led to the definition of three core content pillars.

The first pillar, \textbf{Artificial Intelligence Fundamentals}, covered concepts related to language models, including selection criteria and trade-offs across dimensions such as cost, sustainability, and governance.

The second pillar, \textbf{Domain Contextualization}, was grounded in the application of Domain-Driven Design principles \cite{evans2003} to the context of AI agents, aiming to unify communication between technology and product teams and to ensure that agents accurately reflect domain knowledge and its terminology.

The third pillar, \textbf{Tools and Practice}, emphasized hands-on experience through practical scenarios simulating the day-to-day activities of teams using AI-assisted coding tools.

In addition, the program incorporated two cross-cutting elements: \textbf{mindset}, preparing Advocates to address organizational resistance and adopt a lifelong learning approach; and \textbf{value measurement}, introducing metrics designed to capture impact beyond development speed. A continuous evolution roadmap was also defined to ensure the program remains up to date over time.

\section{Data Collection and Analysis}
To understand participants’ perceptions of their experience in the program, data were collected at two distinct points throughout the journey.

The first instrument was a knowledge assessment administered both before and after the training. Its purpose was to evaluate the retention of concepts covered in the sessions and to enable a comparative analysis of participants’ technical proficiency across the program topics.

The second instrument was a post-training satisfaction survey, distributed after the completion of the four sessions. The survey captured perceptions of content quality, instructor effectiveness, session structure, and suggestions for improvement, using a Net Promoter Score (NPS) scale complemented by open-ended questions. A total of 86 responses were collected.

The results from both instruments are presented in Section 4.

\section{Results}

\subsection{Learning Outcomes}
To evaluate the effectiveness of the training, a knowledge assessment was conducted before and after the program. The comparative analysis demonstrates a clear improvement in learning outcomes. The proportion of participants achieving the maximum score increased from 45.8\% (21 individuals) in the pre-training assessment to 61.7\% (50 individuals) in the post-training assessment, representing an approximate 35\% increase.

In addition to improving top performance, the training also elevated the overall technical level of participants with initially lower scores. This resulted in reduced variance and a more homogeneous distribution of knowledge at a higher level. These findings indicate that the training methodology was effective not only in reinforcing existing knowledge but also in bringing a broader portion of participants to a high-performance level.

\subsection{Satisfaction and Feedback}
Overall, the training was received with strong enthusiasm and was perceived as aligned with current developments in the AI landscape. A total of 86 respondents completed the final evaluation, with 80\% of the Net Promoter Score (NPS) falling within the promoter range (scores of 9 and 10), indicating high levels of participant satisfaction.

Participants highlighted the technical quality and instructional delivery of the speakers as key strengths. The ability to navigate between high-level concepts and practical examples contributed to a more engaging learning experience. Specific elements, such as the “squad transformation pyramid” and the discussion around the orchestrator role within hybrid teams, were perceived as particularly insightful and relevant for professional growth.

From a structural perspective, the division into smaller, sequential sessions was positively received, as it respected participants’ schedules and helped reduce cognitive fatigue. The hands-on format and demonstrations of market-available tools were identified as the moments of highest engagement, enabling participants to connect theoretical concepts with real-world engineering and product scenarios.

Despite the overall positive reception, participants consistently indicated the need for a stronger balance between theory and practice. A recurring suggestion was the inclusion of more hands-on activities, such as real-world workflow examples, code-level demonstrations, efficiency metrics, and case studies reflecting existing squad challenges rather than predominantly greenfield scenarios.

Looking ahead, participants expressed interest in continued learning opportunities with a more practical and specialized focus, enabling the direct application of acquired knowledge in client-facing contexts.

\section{Discussion}
The AI Advocates program provided important insights into the challenges of enabling software engineering professionals to collaborate effectively with human-AI systems.

Programs like this are particularly relevant when considering global trends. Research indicates that collaborative integration between humans and Artificial Intelligence (AI) can increase organizational productivity by up to 40\% \cite{dellacqua2026}. However, these
gains are not automatic; they require structured training and continuous upskilling to ensure individuals are prepared for these changes.

As presented in Section 2.2, the program was originally designed for 40 participants. However, visibility of the syllabus across different leadership levels expanded the audience to more than six times its initial size. This growth was not driven solely by hierarchical mandate, but rather by the perceived relevance of the topic, indicating that AI-driven transformation is recognized as urgent among technical leadership.

The participant profile, concentrated at Senior and Specialist/Tech Lead levels (Section 2.2), was intentional. The program targeted professionals with a multiplier profile, as well as those with the potential to generate greater financial return for the organization through the impact and value articulation of the AI Advocate role in client engagements. However, this concentration raises questions about the effectiveness of the same format for early-career professionals, who may require differentiated learning paths.

The assessment results (Section 4.1) indicate effectiveness in consolidating learning, as evidenced by the increase in the proportion of participants achieving maximum scores from 45.8\% to 61.7\%. However, the instrument primarily measured conceptual retention. Whether
this knowledge translates into daily practices—such as the adoption of Spec-Driven Development or the integration of agents into workflows—remains an open question and justifies longitudinal investigation.

Satisfaction data (Section 4.2) reinforces the program’s positive reception. Participants highlighted the instructional quality and the sequential structure of the sessions as key strengths. However, the most recurring feedback was the desire for more hands-on practice. With six hours of content delivered over three weeks, the trade-off between breadth and depth was inevitable. Prioritizing strategic framing was a deliberate choice, but the feedback
suggests that future initiatives should place greater emphasis on learning grounded in real-world scenarios.

The decision to rely on internal talent proved to be both a practical necessity and a strategic advantage. Instructors with direct experience in the tools and client contexts brought a level of credibility that an external partner would have been unlikely to replicate within the same timeframe.

The emphasis on mindset as a cross-cutting element also deserves attention. Preparing Advocates to navigate organizational resistance—through a posture of knowledge sharing and continuous learning, rather than relying solely on technical arguments—reflects an understanding that AI adoption is as much a change management challenge as it is a technical one.

Finally, it is important to acknowledge the study’s limitations. The data captures short-term perceptions and does not yet demonstrate sustained behavioral change. The program was conducted within a single organization, and the three-week timeframe may have limited the depth of the instructional design. However, this limitation was partially addressed through program expansion via recorded sessions, integration with the company’s technical authority area, and the creation of support groups for knowledge exchange and presentations of practical, day-to-day applications.

\section{Conclusion}
This paper reports on the design and execution of the AI Advocates program at Zup Innovation, aimed at preparing software engineering professionals for the transition to hybrid squads.

The program reached 238 participants across four six-hour sessions, structured around three pillars (AI Fundamentals, Domain Contextualization, Tools and Practice) and two cross-cutting elements (mindset and value measurement). The data collected, detailed in Section 4, indicates positive results in knowledge retention and participant satisfaction, which served as the basis for scaling the initiative through its inclusion in a broader learning track available to the entire organization.

As future work, we intend to conduct a longitudinal analysis of the program’s impact on daily practices, including the adoption of Spec-Driven Development (SDD) and the integration of agents into squad workflows. A second direction involves designing differentiated learning
paths based on seniority, addressing the feedback requesting more applied content. The program’s evolution roadmap includes periodic updates to keep pace with advancements in tools and methodologies.

The experience suggests that the transition to hybrid squads is as much a cultural and educational challenge as it is a technical one. Programs that combine technical content, strategic framing, and Responsible AI principles can accelerate this transition in a productive and ethically grounded way, focusing not only on the intensity of the content, but also on the consistency of knowledge dissemination—whether through recorded sessions or dedicated spaces for experience sharing.

\section{References}
\nocite{wef2023}
\printbibliography

\end{document}